\documentclass[aps,pra,twocolumn,amsmath,amssymb,nofootinbib,showpacs,superscriptaddress]{revtex4-1}
\usepackage[english]{babel}
\usepackage{latexsym}
\usepackage{graphics}
\usepackage{graphicx}
\usepackage{epsfig}
\usepackage{color}
\usepackage{bm}
\usepackage{amsmath}
\usepackage{amssymb}
\usepackage{amsthm}
\usepackage{dcolumn}
\usepackage{bm}
\usepackage{float}
\usepackage{hyperref}
\usepackage{color}
\usepackage{epstopdf}
\usepackage{cleveref}
\usepackage[svgnames]{xcolor}
\hypersetup{hidelinks,colorlinks=true,allcolors=DarkBlue}
\begin{document}

\title{$P$-wave superfluidity of atomic lattice fermions}

\author{A.K. Fedorov}
\affiliation{Russian Quantum Center, Skolkovo, Moscow 143025, Russia}
\affiliation{LPTMS, CNRS, Univ. Paris-Sud, Universit\'e Paris-Saclay, Orsay 91405, France}
\affiliation{Russian Quantum Center, National University of Science and Technology MISIS, Moscow 119049, Russia}

\author{V.I. Yudson}
\affiliation{Laboratory for Condensed Matter Physics, National Research University Higher School of Economics, Moscow 101000, Russia}
\affiliation{Russian Quantum Center, Skolkovo, Moscow 143025, Russia}

\author{G.V. Shlyapnikov}
\affiliation{Russian Quantum Center, Skolkovo, Moscow 143025, Russia}
\affiliation{LPTMS, CNRS, Univ. Paris-Sud, Universit\'e Paris-Saclay, Orsay 91405, France}
\affiliation{Russian Quantum Center, National University of Science and Technology MISIS, Moscow 119049, Russia}
\affiliation{SPEC, CEA, CNRS, Universit\'e Paris-Saclay, CEA Saclay, Gif sur Yvette 91191, France}
\affiliation{Van der Waals-Zeeman Institute, Institute of Physics, University of Amsterdam, Science Park 904, 1098 XH Amsterdam, The Netherlands}
\affiliation{Wuhan Institute of Physics and Mathematics, Chinese Academy of Sciences, 430071 Wuhan, China}

\date{\today}
\begin{abstract}
We discuss the emergence of $p$-wave superfluidity of identical atomic fermions in a two-dimensional optical lattice.
The optical lattice potential manifests itself in an interplay between an increase in the density of states on the Fermi surface 
and the modification of the fermion-fermion interaction (scattering) amplitude.
The density of states is enhanced due to an increase of the effective mass of atoms. 
In deep lattices the scattering amplitude is strongly reduced compared to free space due to a small overlap of wavefunctions of fermions sitting in the neighboring lattice sites, 
which suppresses the $p$-wave superfluidity.
However, for moderate lattice depths the enhancement of the density of states can compensate the decrease of the scattering amplitude. 
Moreover, the lattice setup significantly reduces inelastic collisional losses, which allows one to get closer to a $p$-wave Feshbach resonance.
This opens possibilities to obtain the topological $p_x+ip_y$ superfluid phase, especially in the recently proposed subwavelength lattices. 
We demonstrate this for the two-dimensional version of the Kronig-Penney model allowing a transparent physical analysis.

\begin{description}
\item[PACS numbers]
67.85.De, 03.65.Vf, 03.67.Lx, 03.75.Ss
\end{description}
\end{abstract}
\maketitle

\section{Introduction}

$P$-wave pairing of fermions is a basis of superfluidity in $^{3}$He~\cite{Vollhardt},
and it provides superconductivity in unconventional superconductors~\cite{Mineev}.
Presently, the $p$-wave superfluid pairing attracts a great deal of interest in ultracold atomic gases \cite{Gurarie,Efremov,Gurarie2,Zhang,Sato,Nishida,Bulgac}.
One of the reasons is the search for topological $p_x+ip_y$ superfluid of identical fermions in the two-dimensional (2D) geometry.\,Topological properties of this 
phase emerge from zero-energy Majorana modes on the vortex cores~\cite{Stern}, and
Non-Abelian statistics of the vortices forms a basis for the implementation of topologically protected quantum information processing~\cite{Nayak,Ivanov,Kitaev,Tewari,Sau}.

Despite a significant progress in theory~\cite{Gurarie,Efremov,Gurarie2,Zhang,Sato,Nishida,Bulgac},\,the $p_x+ip_y$ superfluid has not been observed.
The crucial obstacle to achieve this phase for spinless short-range interacting fermions comes from a small value of the $p$-wave interaction.
Therefore, in order to obtain a sizable transition temperature one has to approach a $p$-wave Feshbach resonance.
The $p$-wave resonances have been studied in experiments with fermionic potassium~\cite{Jin,Jin2,Esslinger} and lithium~\cite{Salomon,Ketterle3,Salomon2,Ticknor,Ueda,Nakasuji} atoms.
Close to the resonance the rate of inelastic collisional losses becomes very large~\cite{3body,3body2,3body3}.
Thus, the superfluid of short-range interacting atomic fermions is characterized either by
vanishingly low critical temperature or by instability due to collisional losses.

\begin{figure}[t]
\begin{centering}
\includegraphics[width=0.865\columnwidth]{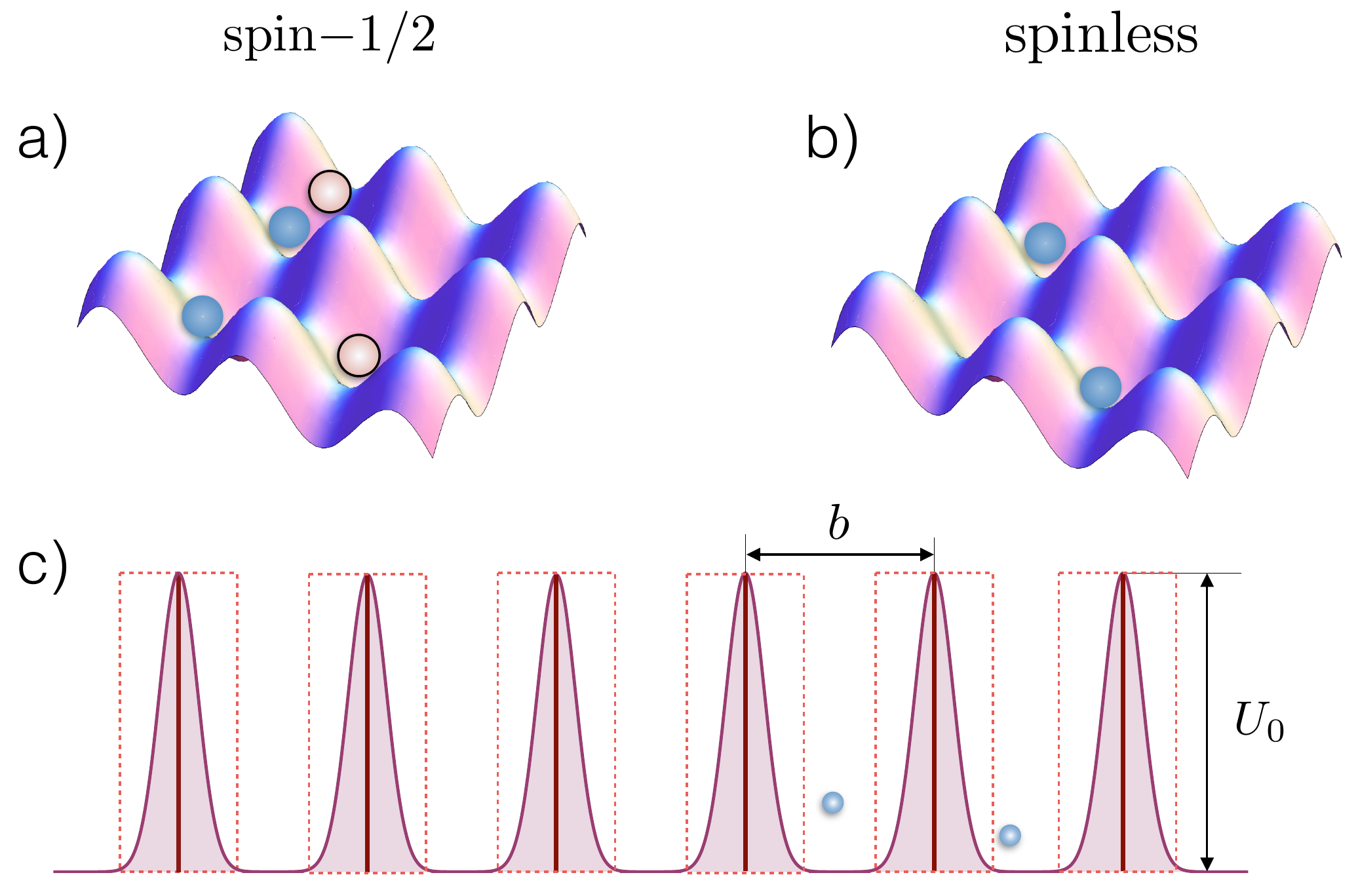}
\end{centering}
\vskip -4mm
\caption
{
Superfluid pairing of lattice fermions in various setups.
In (a) two component (spin-$1/2$) lattice fermions with a short-range interaction.
The two spin components are labeled by filled and unfilled circles.
In (b) single component (spinless) short-range interacting lattice fermions.
In (c) 1D projection of atomic fermions loaded in the 2D Kronig-Penney lattice.
}
\label{fig:setup}
\end{figure}

The creation of $p_x+ip_y$ atomic or molecular topological superfluids in 2D optical lattices can be a promising path for future prospects,
since addressing qubits in the lattice should be much easier than in the gas phase.
For microwave-dressed polar molecules the long-range character of the acquired attractive dipole-dipole intermolecular interaction~\cite{Shlyapnikov,Shlyapnikov2} leads
to similar results regarding the critical temperature as in free space~\cite{Fedorov},
at least in subwavelength lattices.
For short-range interacting atomic fermions the situation is different.
The effect of the lattice potential on the formation of a superfluid phase of atomic fermions has been actively discussed~\cite{Lukin,Lukin2,Lewenstein,Iskin,Ketterle,Zoller,TBM1,TBM2}.
In particular, for the $s$-wave pairing of spin-1/2 fermions an increase in the depth of the optical potential results in a stronger atom localization and hence in increasing the on-site interaction.
At the same time, the tunneling becomes weaker.
The combined effect of these two factors is a strong increase in the critical temperature~\cite{Lukin,Lukin2,Iskin}.
This has been observed in the MIT experiment~\cite{Ketterle}.
For the lattice filling somewhat smaller than unity, the physical picture can be rephrased as follows.
An increase in the lattice depth increases an effective mass of atoms and, hence, makes the density of states (DOS) larger.
The effective fermion-fermion scattering amplitude is also increasing.
The critical temperature in the BCS approach is $T_c\propto\exp\left[-1/\lambda_c\right]$,
where $\lambda_c$ is proportional to the product of the (modulus of) the scattering amplitude and the DOS on the Fermi surface.
Thus, an increase in the lattice potential increases $T_c$.

On the contrary, for identical fermions in fairly deep lattices (tight-binding model) the fermion-fermion scattering amplitude is strongly reduced.
In the lowest band approach two fermions do not occupy the same lattice site,
and the amplitude is proportional to a very small overlap of the wavefunctions of fermions sitting in the neighboring sites.
This suppresses the $p$-wave superfluid pairing for fairly small filling factors in deep lattices, which is consistent with numerical calculations of Ref.~\cite{Iskin}.
Nevertheless, there remains a question about an interplay between an increase of the DOS and the modification of the fermion-fermion scattering amplitude for moderate lattice depths.
However, in sinusoidal optical lattices single particle states are described by complicated Mathieu functions, which complicates the question.

In this paper we study identical fermionic atoms in a 2D version of the Kronig-Penney model allowing a transparent physical analysis for moderate lattice depths.
The 2D version of the Kronig-Penney model is a superposition of two Kronig-Penney potentials (in the $x$ and $y$ directions, respectively).
With the eigenfunctions being piecewise plane waves,
the Kronig-Penney potential is used in cold atom theory (see, e.g.,~\cite{DW2007,Calarco,Lacki2016}) to mimic sinusoidal potentials of common optical lattices.
In particular, this model allows us to investigate two important questions.
The first one is about an interplay between an increase of the DOS and the modification of the fermion-fermion interaction in lattices of moderate depths.
We demonstrate that the reduction of the scattering amplitude still dominates over the enhancement of the DOS.
The second question is about the stability of the system with respect to collisional losses.
We show that the lattice setup reduces inelastic collisional losses compared to free space,
and one can approach the Feshbach resonance without a strong collisional instability.
This opens a possibility to observe the lattice $p_x+ip_y$ 2D superfluid and maybe other interesting many-body phases.

The paper is organized as follows.
In Sec.~\ref{sec:approach} we describe a general approach for studying superfluidity of 2D lattice fermions (Fig.~\ref{fig:setup}).
Sec.~\ref{sec:short} contains the demonstration of how the ordinary tight-binding optical lattice promotes the $s$-wave superfluidity of spin-1/2 fermionic atoms
and suppresses the $p$-wave superfluidity of spinless fermions.
In Sec.~\ref{sec:KP} we develop a theory of $p$-wave superfluidity of spinless fermions in the 2D Kronig-Penney lattice.
In Sec.~\ref{sec:inelastic} we discuss inelastic decay processes in the lattice and in Sec.~\ref{sec:conclusion} we conclude.

\section{General relations}\label{sec:approach}

Let us first present a general framework for the investigation of superfluid pairing of weakly interacting lattice fermions.
We will do this for 2D identical (spinless) fermions, having in mind that the approach for spin-1/2 fermions is very similar.
The grand-canonical Hamiltonian of the system is $\hat{\mathcal{H}}=\hat{H}_{0} +\hat{H}_{\mathrm{int}}$,
and the single particle part is given by [hereinafter we put $\hbar=1$ and set the normalization volume (surface) equal to unity]:
\begin{equation}\label{hamiltonian0}
	\hat{H}_{0}={\int}d^2{\bf r}\,\hat\psi^{\dag}({\bf r}\,)\left[-\frac{\mathbf{\nabla}^2}{2m} + U({\bf r})-\mu\right]\hat\psi({\bf r}),
\end{equation}
with $\mu$ being the chemical potential,
$m$ the particle mass,
$U({\bf r})$ the 2D periodic lattice potential,
and $\hat\psi({\bf r})$ the fermionic field operator.

The term $\hat H_{\mathrm{int}}$ describes the interaction between particles:
\begin{equation}\label{hamiltonianint}
	\hat{H}_{\mathrm{int}}=\frac{1}{2}\int{d^2 r d^2 r'\,\hat\psi^{\dag}({\bf r})\hat\psi^{\dag}({\bf r}')V({\bf r}-{\bf r}')\hat\psi({\bf r}')\hat\psi({\bf r})},
\end{equation}
where $V({\bf r}-{\bf r}')$ is the potential of interparticle interaction of radius $r_0$.

In the absence of interactions,
fermions in the periodic potential $U({\bf r})$ fill single particle energy levels $\varepsilon_{\nu}({\bf k})$ determined by the Schr\"{o}dinger equation:
\begin{eqnarray}\label{Single-particle}
	\left[-\frac{\mathbf{\nabla}^2}{2m}+U({\bf r}) \right]\chi_{\nu {\bf k}}({\bf r})=\varepsilon_{\nu}({\bf k})\chi_{\nu {\bf k}}({\bf r}).
\end{eqnarray}
Here $\nu = 0,1,2, \ldots$ numerates energy bands, the wave vector ${\bf k}=\{k_x,k_y\}$ takes values within the Brillouin zone: $\{-\pi/b{<}k_i{<}\pi/b; i=x,y\}$, and $b$ is the lattice period.
The eigenfunctions $\chi_{\nu {\bf k}}({\bf r})$ obey the periodicity condition
\begin{eqnarray}\label{chi-period}
	\chi_{\nu {\bf k}}({\bf r} + {\bf R}_n) =
	\chi_{\nu {\bf k}}({\bf r})\exp{[i{\bf k}{\bf R}_n]},
\end{eqnarray}
where $n = (n_x, n_y)$ is the index of the lattice site, with integer $n_x, n_y$.
In the described Bloch basis the field operator reads:
\begin{eqnarray}\label{Psi-expansion}
	\hat\psi({\bf r}){=}\sum\nolimits_{\nu, {\bf k}} \hat a_{\nu{\bf k}}\chi_{\nu {\bf k}}({\bf r}),
\end{eqnarray}
with $\hat a_{\nu{\bf k}}$ being the annihilation operator of fermions with quasimomentum ${\bf k}$ in the energy band $\nu$.

We assume a dilute regime where the 2D density $n$ is such that $nb^2{\lesssim}1$, and
all fermions are in the lowest Brillouin zone (hereinafter we omit the corresponding index $\nu=0$). In the low momentum limit (small filling factor) that we consider, their Fermi
energy $E_F$ is small compared to the energy bandwidth $E_B$.
The lattice potential amplitude $U_0$ is assumed to be sufficiently large, so that
both $E_F$ and $E_B$ are smaller than the gap between the
first and second lattice bands.
The single particle dispersion relation then takes the form:
\begin{equation}\label{effdisp}
	E_k=\frac{k^2}{2m^*},
\end{equation}
where $m^*>m$ is the effective mass.

In 2D the transition of a Fermi gas from the normal to superfluid state is set by the Kosterlitz-Thouless mechanism. However, in the weakly interacting regime the Kosterlitz-Thouless transition  temperature is very close to $T_c$ calculated in the Bardeen-Cooper-Schrieffer (BCS) approach \cite{Miyake}.
We then reduce the Hamiltonian given by Eqs. (\ref{hamiltonian0}) and (\ref{hamiltonianint}) to the standard BCS form:
\begin{equation}\label{hamiltonian-bcs}
	\begin{aligned}
		\hat{\mathcal{H}}_{\rm BCS}=\sum\nolimits_{{\bf k}}&\left\{(E_k-\mu)\hat{a}^{\dag}_{ {\bf k}}\hat{a}^{}_{{\bf k}}\right. \\
		&\left.+\frac{1}{2}\left[\hat a^{\dag}_{{\bf k}}\hat a^{\dag}_{-{\bf k}}\Delta({\bf k}) + {\rm h.c.}\right]\right\},
	\end{aligned}
\end{equation}
where the momentum-space order parameter $\Delta({\bf k})$ is given by
\begin{equation}\label{Delta-definition}
	\Delta({\bf k})=\sum\nolimits_{{\bf k}\,'}V({\bf k},{\bf k}')\langle \hat a_{-{\bf k}'} \hat a_{{\bf k}'}\rangle,
	\,\,\,
	\Delta({\bf k})=-\Delta(-{\bf k}),
\end{equation}
with $V({\bf k},{\bf k}')$ being the matrix element of the interaction potential between the corresponding states.

The Hamiltonian (\ref{hamiltonian-bcs}) is then decomposed in a set of independent quadratic Hamiltonians
and the anomalous averages are determined by the standard BCS expressions:
\begin{eqnarray} \label{BCS-average}
	\langle{\hat a_{-{\bf k}} \hat a_{{\bf k}}}\rangle=-\Delta({\bf k})\mathcal{K}(k),
\end{eqnarray}
where $\mathcal{K}(k)=\tanh[\mathcal{E}(k)/2T]/2\mathcal{E}(k)$, 
and
\begin{eqnarray} \label{BCS energy}
	\mathcal{E}(k)=\sqrt{(E_k-\mu)^2 + |\Delta_{\nu}(k)|^2}
\end{eqnarray}
is the energy of excitation with quasimomentum ${\bf k}$.
From Eqs.~(\ref{Delta-definition}) and (\ref{BCS-average}) we have an equation
for $\Delta({\bf k})$ (gap equation):
\begin{equation}\label{Delta-equation}
	\Delta({\bf k})=-\sum\nolimits_{{\bf k}'}V({\bf k},{\bf k}')\mathcal{K}(k')\Delta({\bf k}').
\end{equation}
Eq.~(\ref{Delta-equation}) can be expressed \cite{Shlyapnikov2} in terms 
of the effective off-shell scattering amplitude $f({\bf k}',{\bf k})$ of a fermion pair with momenta ${\bf k}$ and $-{\bf k}$ defined as
\begin{equation}\label{scattering}
	\begin{aligned}
		 f({\bf k}',{\bf k})=\int d^2r_{1} d^2r_{2}\,&\Phi^{(0)*}_{{\bf k}'}({\bf r}_1,{\bf r}_2)\\
	 	&\times V({\bf r}_1-{\bf r}_2)\Phi_{{\bf k}}({\bf r}_1,{\bf r}_2).
	\end{aligned}
\end{equation}
Here
\begin{eqnarray} \label{Phi-0}
	\Phi^{(0)}_{{\bf k}}({\bf r}_1, {\bf r}_2)=\chi_{{\bf k}}({\bf r}_1)\chi_{-{\bf k}}({\bf r}_2),
\end{eqnarray}
is the wavefunction of a pair of non-interacting fermions with quasimomenta ${\bf k}$ and $-{\bf k}$.
The quantity $\Phi_{{\bf k}}({\bf r}_1,{\bf r}_2)$ is the true (i.e., accounting for the interaction) wavefunction,
which develops from the incident wavefunction $\Phi^{(0)}_{{\bf k}}({\bf r}_1,{\bf r}_2)$ of a free pair.
The wavefunction $\Phi_{{\bf k}}({\bf r}_1,{\bf r}_2)$ satisfies the Schr\"{o}dinger equation 
\begin{equation}
[\hat{H}_{12}-2E_k]\Phi_{{\bf k}}({\bf r}_1,{\bf r}_2)=0, 
\end{equation}
with the two-particle Hamiltonian:
\begin{eqnarray} \label{Phi-equation}
	\!\!\!\hat{H}_{12}=-\frac{\mathbf{\nabla}^2_1+\mathbf{\nabla}^2_2}{2m}+U({\bf r}_1)+U({\bf r}_2)+V({\bf r}_1{-}{\bf r}_2).
\end{eqnarray}
The renormalized gap equation for the function $\Delta({\bf k})$ then takes the form similar to that in free space (see Ref.~\cite{Shlyapnikov2} and references therein):
\begin{equation}\label{eq:gap}
\begin{split}
	\Delta({\bf k})=\int\frac{d^2k'}{(2\pi)^2}&f({\bf k}',{\bf k})\Delta({\bf k}')\\
	&\times\left\{\mathcal{K}(k')-\frac{1}{2(E_{k'}-E_k)}\right\}.
\end{split}
\end{equation}
In the weakly interacting regime the chemical potential coincides with the Fermi energy $E_F= k_F^2/2m^*$, where $k_F=\sqrt{4\pi{n}}$ is the Fermi momentum.
Note that we omit a correction to the bare interparticle interaction due to polarization of the medium by colliding particles \cite{GMB}.

We will see below that the scattering amplitude and the corresponding critical temperature of the superfluid transition of lattice fermions depend drastically on the presence or absence of spin
and on the pairing angular momentum.
Before analyzing various regimes, we discuss the situation in general.

The efficiency of superfluid pairing first of all depends on the symmetry of the order parameter.
For the pairing with orbital angular momentum $l$ we have $\Delta({\bf k})\rightarrow\Delta_l(k)\exp\left[il\phi_{\bf k}\right]$,
where $\phi_{\bf k}$ is the angle of the vector ${\bf k}$ with respect to the quantization axis.
Integrating Eq.~(\ref{eq:gap}) over $\phi_{\bf k}$ and $\phi_{{\bf k}'}$ we obtain the same equation in which $\Delta({\bf k})$ and $\Delta({\bf k}')$ are replaced with $\Delta_l(k)$ and $\Delta_l(k')$,
and $f({\bf k}',{\bf k})$ is replaced with its $l$-wave part
\begin{equation}\label{fl}
	f_l(k',k)=\int \frac{d\phi_{\bf k}d\phi_{{\bf k}'}}{(2\pi)^2}f({\bf k}',{\bf k})\exp\left[il\phi_{\bf k}-il\phi_{{\bf k}'}\right].
\end{equation}
Alternatively, we can write
\begin{equation}\label{flalt}
	\begin{aligned}
		f_l(k',k)=\int d^2r_1&d^2r_2\Phi^{(0)*}_{lk'}({\bf r}_1,{\bf r}_2)\\
	 	&\times{V(|{\bf r}_1-{\bf r}_2|)\Phi_{lk}({\bf r}_1,{\bf r}_2)}.
	\end{aligned}
\end{equation}
where the $l$-wave parts of the wavefunctions, $\Phi^{(0)}_{lk'}$ and $\Phi_{lk}$, are given by
\begin{eqnarray}
	&&\Phi^{(0)}_{lk'}({\bf r}_1,{\bf r}_2)=\int\frac{d\phi_{{\bf k}'}}{2\pi}\Phi^{(0)}_{{\bf k}'}({\bf r}_1,{\bf r}_2)\exp\left[il\phi_{{\bf k}'}\right],
	\label{Phi0l} \\
	&&\Phi_{lk}({\bf r}_1,{\bf r}_2)=\int\frac{d\phi_{{\bf k}}}{2\pi}\Phi_{{\bf k}}({\bf r}_1,{\bf r}_2)\exp\left[il\phi_{{\bf k}}\right].
	\label{Phil}
\end{eqnarray}
As well as in free space (see Ref.~\cite{Shlyapnikov2}), we turn from $f_l(k',k)$ to the (real) function
\begin{equation}
	\tilde f_l(k',k)=f_l(k',k)\left[1-i\tan\delta(k)\right],
\end{equation}
where $\delta(k)$ is the scattering phase shift. This leads to the gap equation:
\begin{equation}\label{eq:gapk}
\begin{split}
	\Delta_l(k)=-P\int\frac{d^2k'}{(2\pi)^2}&\tilde{f}_l(k',k)\Delta_l(k')\\
	&\times\left\{\mathcal{K}(k')-\frac{1}{E_{k'}-E_k}\right\},
\end{split}
\end{equation}
where the symbol $P$ denotes the principal value of the integral.

In order to estimate the critical temperature $T_c$, we first put $k=k_F$ and notice that the main contribution to the integral over $k'$ in Eq.~(\ref{eq:gapk}) comes from $k'$ close to $k_F$.
At temperatures $T$ tending to the critical temperature $T_c$ from below, we put $\mathcal{E}(k')=|E_{k'}-E_F|$ in $\mathcal{K}(k')$.
Then for the pairing channel related to the interaction with orbital angular momentum $l$, we have the following estimate:
\begin{equation}\label{eq:Tc0}
	T_c\sim E_F\exp\left[-\frac{1}{\lambda_c}\right],
	\quad
	\lambda_c=\rho(k_F)|f_l(k_F)|.
\end{equation}
The quantity $\rho(k_F){=}m^*/2\pi$ is the effective density of states on the Fermi surface,
and $f_l(k_F)$ is the off-shell $l$-wave scattering amplitude of lattice fermions.
The derivation for spin-1/2 fermions with attractive intercomponent interaction leads to the same gap equations (\ref{eq:gap}), (\ref{eq:gapk}) and estimate (\ref{eq:Tc0}) in which
\begin{equation}
	\Delta({\bf k})=\sum_{{\bf k}'}V({\bf k},{\bf k}')\langle\hat a_{\downarrow -{\bf k}'}\hat a_{\uparrow {\bf k}'}\rangle
\end{equation}
and $f({\bf k}',{\bf k})$, $f_l(k',k)$ are the amplitudes of the intercomponent interaction.

Eq.~(\ref{eq:Tc0}) shows that compared to free space we have an additional pre-exponential factor $m/m^*<1$. 
Assuming that the lattice amplitude $f_l(k_F)$ and the free-space amplitude $f_l^0(k_F)$ are related to each other as
\begin{equation}\label{f-lat}
	f_{l}(k_F)=\mathcal{R}_{l}f_l^0(k_F),
\end{equation}
we see that the exponential factor $\lambda_c$ in Eq.~(\ref{eq:Tc0}) becomes
\begin{equation}\label{eq:lambda}
	\lambda_{c}=\mathcal{R}_{l}\frac{m^*}{m}\lambda^0_c,
\end{equation}
where $1/\lambda^0_c$ is the BCS exponent in free space.
Below we compare $T_c$ in various lattice setups with the critical temperature in free space.

\section{Short-range interacting fermionic atoms in a deep 2D lattice}\label{sec:short}

We start with the analysis of superfluid pairing in deep 2D lattices. As an example, we consider a quadratic lattice with the lattice potential of the form:
\begin{eqnarray}\label{eq:U}
	U({\bf r})=U_0\left[\cos{\left(\frac{2\pi}{b}x\right)}+\cos{\left(\frac{2\pi}{b}y\right)}\right].
\end{eqnarray}
For sufficiently deep lattices, the single particle wavefunction has the Wannier form:
\begin{equation}\label{eq:wavefunction}
	\chi_{{\bf k}}({\bf r})=\frac{1}{\sqrt{\mathcal{N}}}\sum_j\phi_0({\bf r}-{\bf R}_j)\exp[i{\bf k}{\bf R}_j],
\end{equation}
where $\mathcal{N}$ is the number of lattice sites. The ground state wavefunction in the lattice cell has an extention $\xi_0$ and is given by
\begin{equation}\label{eq:wavefunctionground}
	\phi_0({\bf r})=\frac{1}{\sqrt{\pi}\xi_0}\exp\left[-\frac{r^2}{2\xi_0^2}\right].
\end{equation}

Using a general formula for the effective mass from Ref.~\cite{LL9},
for a deep potential of the form (\ref{eq:U}) one obtains:
\begin{equation}\label{eq:mass}
	\frac{m^*}{m}\simeq\pi\frac{\xi_0^2}{b^2}\exp\left[\frac{2}{\pi^2}\frac{b^2}{\xi_0^2}\right].
\end{equation}

We will consider fermionic atoms interacting with each other via a short-range potential $V({\bf r})$ of radius $r_0$ and assume the following hierarchy of length scales:
\begin{equation}\label{eq:hierarchy}
	r_0\ll {\xi_{0}}<{b}<{1/k_{F}}.
\end{equation}
We first discuss the $s$-wave pairing of spin-1/2 fermions with attractive intercomponent interaction ($l=0$).

Turning to Eq.~(\ref{flalt}) for $l=0$,
we notice that the main contribution to the $s$-wave scattering amplitude in the lattice comes from the interaction between spin-up and spin-down fermions sitting in one and the same lattice site.
The wavefunctions $\Phi^{(0)}_{0k'}$ and $\Phi_{0k}$ can be written as
\begin{eqnarray}
	&&\Phi_{0k'}^{(0)}({\bf r}_1,{\bf r}_2)=\chi_0({\bf r}_1)\chi_0({\bf r}_2), \\
	&&\Phi_{0k}({\bf r}_1,{\bf r}_2)=\chi_0({\bf r}_1)\chi_0({\bf r}_2)\zeta_0(|{\bf r}_1-{\bf r}_2|),
\end{eqnarray}
where the function $\zeta_0(|{\bf r}_1-{\bf r}_2|)$ is a solution of the Schr\"{o}dinger equation for the $s$-wave relative motion of two particles in free space at zero energy,
and it is tending to unity for interatomic separations greatly exceeding $r_0$.
We put $l=0$ in Eq.~(\ref{flalt}) and integrate over ${\bf r}={\bf r}_1-{\bf r}_2$ and ${\bf r}_+=({\bf r}_1+{\bf r}_2)/2$.
Then, owing to the inequality $r_0\ll\xi_0$, this equation is reduced to
\begin{equation}   \label{f0}
	f_0(k',k)=\int d^2r\,V(r)\zeta(r)\,\int d^2r_+|\chi_0(r_+)|^4.
\end{equation}
Recalling that in the low momentum limit the free space scattering amplitude is given by
\begin{equation}
	f_0^0=\int V(r)\zeta(r)d^2r
\end{equation}
and using Eq.~(\ref{eq:wavefunction}) for the function $\chi_0(r)$, we obtain for the ratio of the lattice to free space amplitude:
\begin{equation}\label{eq:RS}
	\mathcal{R}_{l=0}=\frac{1}{2\pi}\frac{b^2}{\xi_0^2},
\end{equation}
where we made a summation over the lattice sites and put $\mathcal{N}=1/b^2$ as the normalization volume is set to be unity. Thus, according to Eqs. (\ref{eq:lambda}) and (\ref{eq:mass}) the BCS exponent $\lambda_c^{-1}$ becomes smaller than in free space by the following factor:
\begin{equation}\label{eq:enhanceSshort}
	\mathcal{R}_{l=0}\,\frac{m^*}{m}\simeq\frac{1}{2}\exp\left[\frac{2}{\pi^2}\frac{b^2}{\xi_0^2}\right].
\end{equation}

\begin{figure}[h!]
\begin{centering}
\includegraphics[width=0.9\columnwidth]{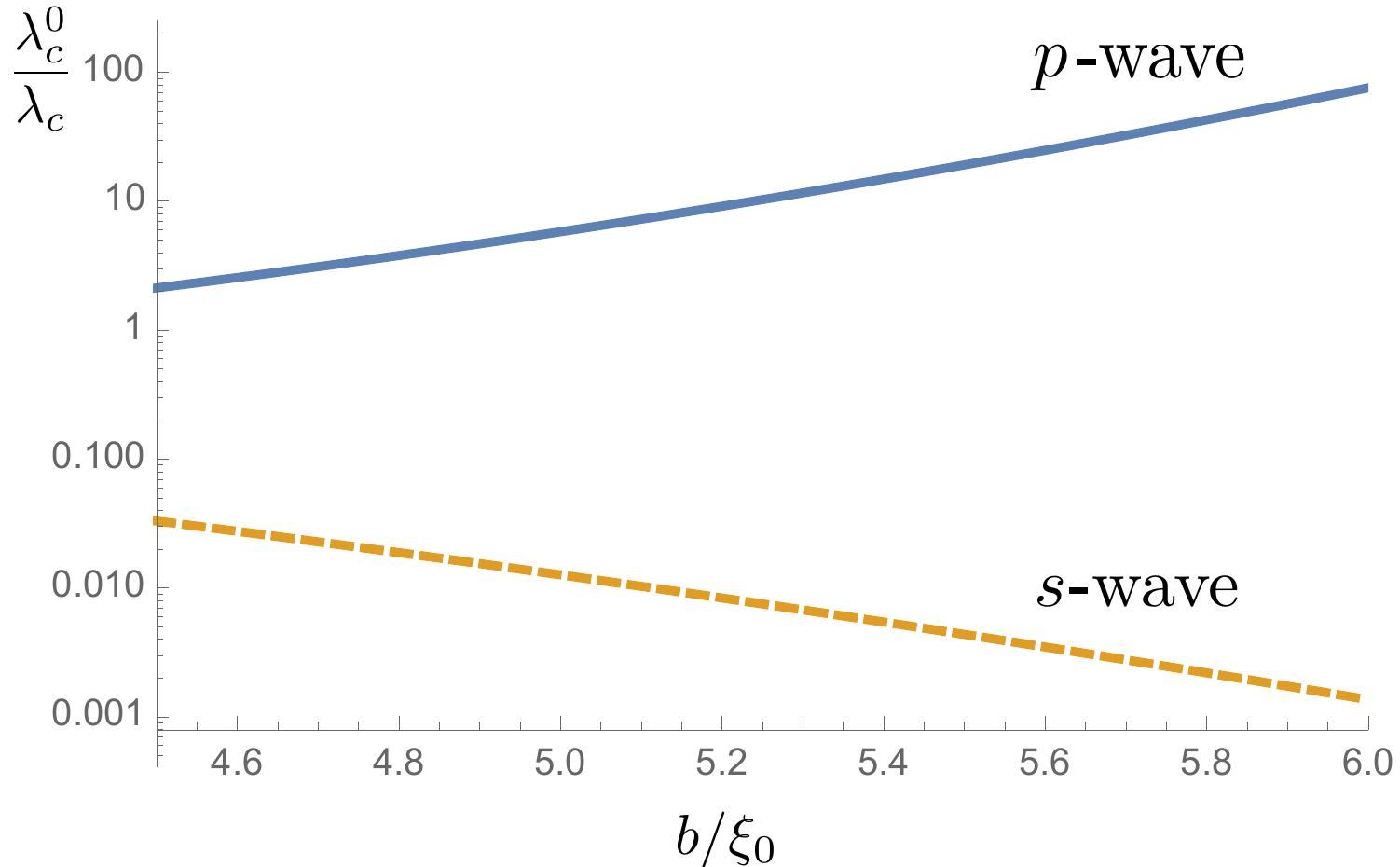}
\end{centering}
\vskip -4mm
\caption
{
The ratio of the BCS exponent in the tight-binding sinusoidal lattice to the BCS exponent in free space, $\lambda_c^0/\lambda_c$, at the same density and short-range coupling strength.
The dashed curve shows $\lambda_c^0/\lambda_c$ as a function of the lattice period (in units of the harmonic oscillator length $\xi_0$) for the $s$-wave pairing of spin-1/2 fermions,
and the solid curve is $\lambda_c^0/\lambda_c$ for the $p$-wave pairing of identical fermions.
}
\label{fig:enhancement}
\end{figure}

For example, taking $b/\xi_0=4$ the BCS exponent $\lambda_c^{-1}$ decreases by a factor of $0.08$,
whereas the effective mass becomes higher by a factor of 5 compared to the bare mass $m$ (see Fig.~\ref{fig:enhancement}).
Then, for $^6$Li atoms at density $10^8$ cm$^{-2}$ ($b\simeq 250$ nm, $k_Fb\simeq 0.5$) we have the Fermi energy $\sim 40$ nK.
Assuming that the free space BCS exponent is about 30 and the related critical temperature is practically zero, in the lattice we obtain $T_c\sim 3$ nK.
We thus see that the lattice setup may strongly promote the $s$-wave superfluidity of spin-1/2 fermions.

The situation with $p$-wave superfluidity of identical fermions is drastically different.
In the single band approximation (tight binding model) two such fermions can not occupy one and the same lattice site.
This is clearly seen using the functions $\chi_{\bf k}({\bf r}_1)$ and $\chi_{-{\bf k}}({\bf r}_2)$ from Eq.~(\ref{eq:wavefunction}) at the same ${\bf R}_j$,
so that the wavefunction $\Phi^{(0)}_{{\bf k}'}({\bf r}_1,{\bf r}_2)$  becomes independent of ${\bf k}'$.
Therefore, the $p$-wave part of this wavefunction $\Phi^{(0)}_{1k'}$ and the $p$-wave scattering amplitude $f_1(k',k)$ following from Eqs.~(\ref{Phi0l}) and (\ref{flalt}) at $l=1$ are equal to zero.

The main contribution to the interaction amplitude then comes from the overlap of the wavefunctions of fermions sitting in the neighboring sites.
We then use Eqs. (\ref{eq:wavefunction}) and (\ref{eq:wavefunctionground}) and write:
\begin{equation}
\begin{split}
&\Phi^{(0)}_{{\bf k}'}({\bf r}_1,{\bf r}_2)=\chi_{{\bf k}'}({\bf r}_1)\chi_{-{\bf k}'}({\bf r}_2)=(1/\mathcal{N}\pi \xi_0^2)\times \\
&\!\!\!\!\!\!\sum_{i,j}\exp\left\{\!-({\bf r}_1\!-\!{\bf R}_i)^2\!/2\xi_0^2\!-\!({\bf r}_2\!-\!{\bf R}_j)^2\!/2\xi_0^2\!-\!i{\bf k}'{\bf b}_j\!\right\}\!,\!  \label{Phi10} 
\end{split}
\end{equation}
with ${\bf b}_j={\bf R}_j-{\bf R}_i$ and ${\bf R}_i,{\bf R}_j$ being the coordinates of the sites $i$ and $j$.  
For the short-range interaction between particles the main contribution to the scattering amplitude comes from distances ${\bf r}_1,{\bf r}_2$ 
that are very close to each other, and for given $i,j$ both coordinates should be close to $({\bf R}_j+{\bf R}_i)/2$. 
Therefore, Eq.~(\ref{Phi10}) is conveniently rewritten as 
\begin{eqnarray}
&\Phi^{(0)}_{{\bf k}'}({\bf r}_1,{\bf r}_2)=(1/\mathcal{N}\pi \xi_0^2)\sum_{i,j}\exp\Big\{-i{\bf k}'{\bf b}_j-r_{+j}^2/\xi_0^2-   \nonumber \\
&r^2/4\xi_0^2-b^2/4\xi_0^2-{\bf r}{\bf b}_j/2\xi_0^2\Big\},    \label{Phi20}
\end{eqnarray}
where ${\bf r}={\bf r}_1-{\bf r}_2$, ${\bf r}_{+j}={\bf r}_+-({\bf R}_i+{\bf R}_j)/2$, ${\bf r}_+=({\bf r}_1+{\bf r}_2)/2$, 
and the summation is performed over the sites $j$ that are nearest neighbours of the site $i$. 
Assuming the conditions $k'b\ll 1$ and $r\sim r_0\ll \xi_0^2/b\ll\xi_0$, for the $p$-wave part of this wavefunction equation (\ref{Phi0l}) at $l=1$ gives:
\begin{equation}
\begin{split} \label{Phi0p} 
\!\!\!\!\!\!\!\!\Phi^{(0)}_{1k'}(r,r_+,\phi_{\bf r})\!&=\!\!\frac{k'rb^2}{\mathcal{N}8\pi\xi_0^4}\!\sum_{i,j}\!\!\exp\!\left\{\!-\!r_{+j}^2/\!\xi_0^2\!-\!\!b^2/\!4\xi_0^2\!\right\}\!\!   \\ 
&\times[\exp(i\phi_{\bf r})+\exp(-i\phi_{\bf r}+2i\phi_j)],   
\end{split}
\end{equation}
where $\phi_{\bf r}$ and $\phi_j$ are the angles of the vectors ${\bf r}$ and $b_j$ with respect to the quantization axis. The $p$-wave part of the true relative-motion wavefunction $\Phi_{\bf k}({\bf r}_1,{\bf r}_2)$ under the same conditions is given by
\begin{equation}
\begin{split}\label{Phip}
	\Phi_{1k}(r,r_+,\phi_{\bf r})&= \frac{b^2}{\mathcal{N}4\pi\xi_0^4}\zeta_1(r)\sum_{i',j'}\exp\left\{\!-\!r_{+j'}^2/\xi_0^2\!-\!b^2/4\xi_0^2\right\} \\
	&\times[\exp(i\phi_{\bf r})+\exp(-i\phi_{\bf r}+2i\phi_j)].
\end{split}
\end{equation}
The function $\zeta_1(r)$ is a solution of the Schr\"{o}dinger equation for the $p$-wave relative motion of two particles at energy tending to zero in free space. 
Sufficiently far from resonance, where the on-shell scattering amplitude satisfies the inequality $m|f_1(k)|\ll 1$, the function $\zeta_1(r)$ becomes $kr/2$ at distances $r\gg r_0$.

Looking at the product of the free and true relative-motion wavefunctions we notice that the main contribution to the scattering amplitude (\ref{flalt}) comes from the terms in which 
${\bf R}_i+{\bf R}_j={\bf R}_{i'}+{\bf R}_{j'}$, i.e. ${\bf r}_{+j}={\bf r}_{+j'}$. 
This is realized for $i=i'$, $j=j'$ or $i'=j$, $j'=i$.
Then, recalling that for $k'r_0\ll 1$ and $kr_0\ll 1$ the free space on-shell scattering amplitude is
\begin{equation}
	f_1^0(k',k)=\int V(r)(k'r/2)\zeta_1(r)d^2r,
\end{equation}
we first integrate each term of the sum over $i,j,i'j'$ in the product $\Phi^{(0)*}_{1k'}\Phi_{1k}$ over $d^2r$ and $d^2r_+$ in Eq.~(\ref{flalt}). 
After that we make a summation over the neighboring sites $j$ and over the sites $i$ and take into account that $\mathcal{N}=1/b^2$. 
Eventually, this gives for the ratio of the lattice to free space $p$-wave amplitude:
\begin{equation}\label{Rl1}
	\mathcal{R}_{l=1}=\frac{1}{2\pi}\left(\frac{b}{\xi_0}\right)^6\exp\left[-\frac{b^2}{2\xi_0^2}\right].
\end{equation}
Thus, with the help of Eq.~(\ref{eq:mass}) the inverse BCS exponent in the lattice becomes:
\begin{equation}\label{lambdac1}
	\lambda_c=\mathcal{R}_{l=1}\frac{m^*}{m}\lambda_c^0=\frac{\lambda_c^0}{2}\left(\frac{b}{\xi_0}\right)^4\exp\left[-\frac{cb^2}{\xi_0^2}\right],
\end{equation}
where $c\simeq{0.3}$.

We now clearly see that the inverse BCS exponent $\lambda_c$ in the lattice is exponentially smaller compared to its value in free space.
In particular, already for $b/\xi_0=5$ the ratio $\lambda_c^0/\lambda_c$ it is about $6$, which practically suppresses $p$-wave superfluidity of identical fermions (see Fig.~\ref{fig:enhancement}).
However, this ratio rapidly reduces with decreasing the ratio $b/\xi_0$ and becomes $\sim 1$ for $b/\xi_0=4$.
It is therefore interesting to analyze more carefully the case of moderate lattice depths.

\section{Superfluid $P$-wave pairing in the 2D Kronig-Penney lattice}\label{sec:KP}

We will do so using a 2D version of the Kronig-Penney model,
namely a superposition of two 1D Kronig-Penney lattices (in the $x$ and $y$ directions, respectively), with a $\delta$-functional form of potential barriers:
\begin{equation}\label{eq:UKP}
	U(x,y)=U_{0}b\sum_{j=-\infty}^{+\infty}\left[\delta(x-jb)+\delta(y-jb)\right].
\end{equation}
With the eigenfunctions being piecewise plane waves,
the 1D Kronig-Penney potential is used in ultracold atom theory (see, e.g. \cite{DW2007,Calarco,Lacki2016}) to mimic sinusoidal potentials.
The model (\ref{eq:UKP}) catches the key physics and allows for transparent calculations.
The latter circumstance is a great advantage compared to sinusoidal lattices where single particle states are described by complicated Mathieu functions.
The considered model allows us to investigate two important questions.
The first question is about an interplay between an increase of the DOS and the modification of the fermion-fermion interaction for moderate lattice depths.
The second one is the stability of the system with respect to collisional losses.

Single-particle energies in the periodic potential (\ref{eq:UKP}) are represented as
\begin{equation}\label{eq:DKP}
	E_{\textbf k}=E(k_x)+E(k_y),
\end{equation}
where $E(k_{x,y})>0$ is the dispersion relation for the 1D Kronig-Penney model.
It follows from the equation (see, e.g., Ref. \cite{LL9}):
\begin{equation}\label{eq:DKP}
	\cos(q{b})+G\frac{\sin(q{b})}{qb}=\cos(kb),
\end{equation}
where $q{=}\sqrt{2mE(k)}>0$, and $G=mU_0b^2$.
As well as in the previous section,
we consider a dilute regime where the filling factor is $\nu=nb^2{\lesssim}1$ and the fermions fill only a small energy interval near the bottom of the lowest Brillouin zone.
Then the energy counted from the bottom of the zone is given by Eq.~(\ref{effdisp}) and for the effective mass Eq.~(\ref{eq:DKP}) yields:
\begin{equation}\label{eq:mKP}	
	\frac{m^*}{m}\approx\frac{\tan(\eta/2)}{\eta}\left[{1+\frac{\sin\eta}{\eta}}\right],
\end{equation}
with $\eta$ being the smallest root of the equation:
\begin{equation}\label{eq:eta}
	\eta\tan(\eta/2)=G.
\end{equation}
Actually, $\eta=q_0b$ where $q_0$ follows from Eq.~(\ref{eq:DKP}) at $k=0$.

For $m^*\gg{m}$ we have $m^*/m=G/\pi^2$, which means that the quantity $G$ should be very large.
Then the width of the lowest Brillouin zone is $E_B=2/m^*b^2$ and it is much larger than the Fermi energy $E_F=k^2/2m^*$ for $k_Fb<0.5$.
The gap between the lowest and second zones is $E_G=3\pi^2/2mb^2$ and it greatly exceeds $E_B$ and $E_F$.
Note that even for $m^*\simeq1.3m$ ($G\simeq 5$) we have $E_G$ close to $4E_B$, and the ratio $E_F/E_B$ is significantly smaller than unity if $k_Fb<0.5$.
This justifies the single-band approximation and the use of the quadratic dispersion relation (\ref{effdisp}).

\begin{figure}[t]
\begin{centering}
\includegraphics[width=0.8\columnwidth]{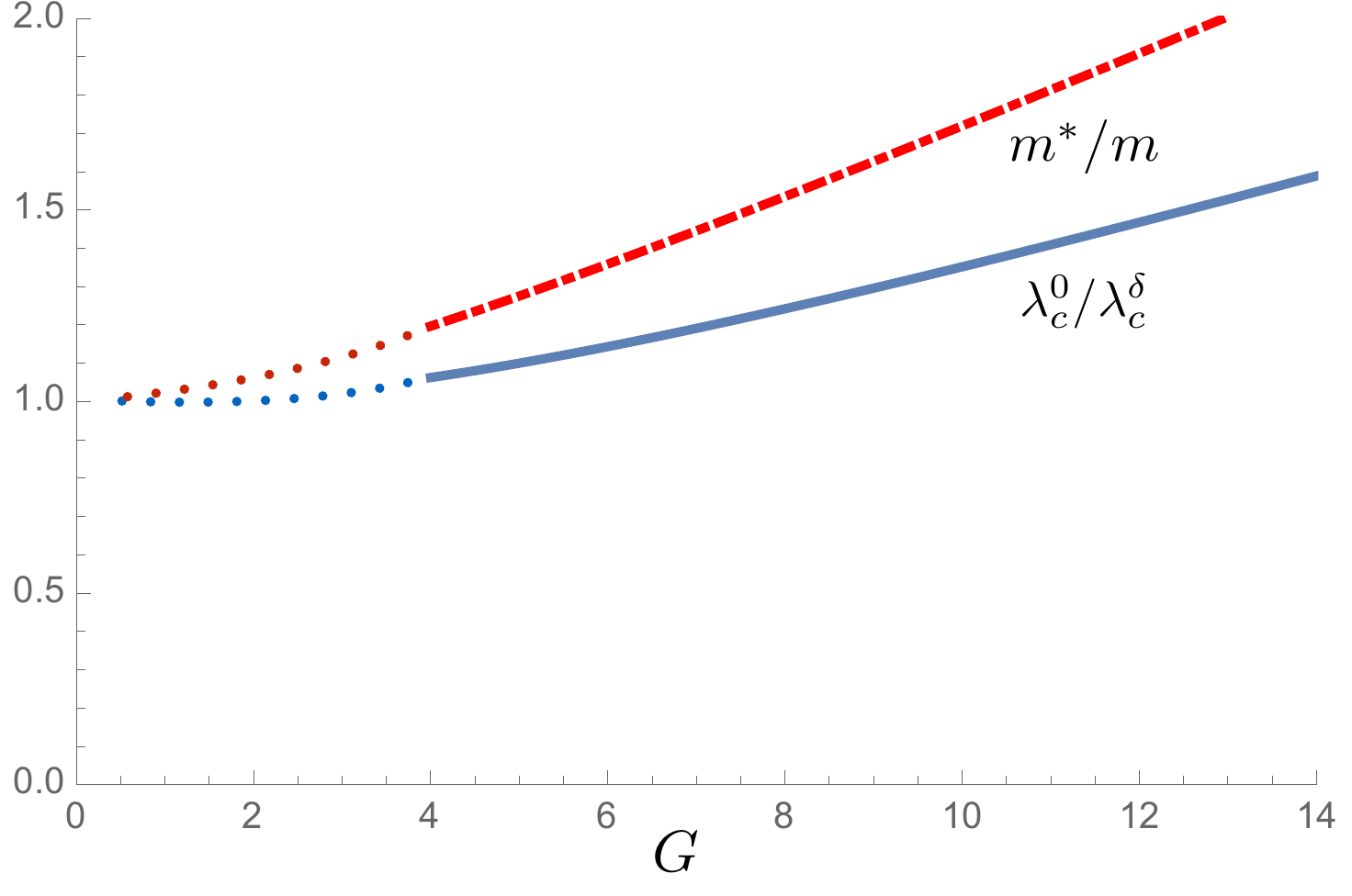}
\end{centering}
\vskip -4mm
\caption
{
The ratio of the BCS exponent in the 2D $\delta$-functional Kronig-Penney lattice to the BCS exponent in free space, $\lambda_c^0/\lambda_c^{\delta}$,
at the same density and short-range coupling strength.
The solid blue curve shows $\lambda_c^0/\lambda_c^{\delta}$ as a function of the lattice depth $G$,
and the dashed red curve the effective mass $m^*/m$ versus $G$. The dotted parts of these curves show our expectation at $G\lesssim 1$, 
where the single-band approximation used in our calculations does not work.
}
\label{fig:KP}
\end{figure}

Single-particle wavefunctions $\chi_{\bf k}(\bf r)$ are of the form $\chi_{\bf k}({\bf r})=\chi_{k_x}(x)\chi_{k_y}(y)$, where
\begin{equation}\label{eq:chi-exact}
\begin{split}
	\!\!\!\!\chi_{k_x}{(x)}{=}&\frac{\sqrt{2}\sin\left(\eta/2\right)}{\sqrt{1+{\sin\eta}/{\eta}}}\sum_{j=-\infty}^{j=+\infty}{\!\!A_j(x)\exp{[ik_xjb]}}\\
	&\times\left\{\frac{e^{iq{b}}e^{iq{(x-jb)}}}{e^{iq{b}}-e^{ik_x{b}}}-\frac{e^{-iq{b}}e^{-iq{(x-jb)}}}{e^{-iq{b}}-e^{ik_x{b}}}\right\}
\end{split}
\end{equation}
is the exact eigenfunction of the 1D Kronig-Penney model, with $A_j(x)=1$ for $(j-1)b<x<jb$ and zero otherwise.
The function $\chi_{k_y}(y)$ has a similar form.
For $k'b\ll 1$ and $kb\ll 1$ the $p$-wave parts of the wavefunctions, $\Phi^{(0)}_{1k'}$ and $\Phi_{1k}$, following from Eqs. (\ref{Phi-0}), (\ref{Phi0l}), and (\ref{Phil}) at $l=1$ turn out to be
\begin{eqnarray}
	&&\Phi^{(0)}_{1k'}=ik'r\frac{\eta\cot(\eta/2)}{[1+\sin\eta/\eta]^2}\sum_{j_x,j_y=-\infty}^{\infty}A_{j_x}(x_+)A_{j_y}(y_+) \label{Phi0KP} \\
	&&\!\!\times\!\!\left\{\!\cos\phi_{\bf r}\!\cos^2\!\!\left(\!\!q_0y_+\!\!-\!j_yb\!+\!\frac{b}{2}\!\right)\!+\!i\sin\phi_{\bf r}\!\cos^2\!\!\left(\!\!q_0x_+\!\!-\!j_xb\!+\!\frac{b}{2}\!\right)\!\!\right\},  \nonumber \\
	&&\Phi_{1k}=2i\zeta_1(r)\frac{\eta\cot(\eta/2)}{[1+\sin\eta/\eta]^2}\sum_{j_x,j_y=-\infty}^{\infty}\!\!\!\!\!A_{j_x}(x_+)A_{j_y}(y_+) \label{PhiKP}\\
	&&\!\!\times\!\!\left\{\!\cos\phi_{\bf r}\!\cos^2\!\!\left(\!\!q_0y_+\!\!-\!j_yb\!+\!\frac{b}{2}\!\right)\!+\!i\sin\phi_{\bf r}\!\cos^2\!\!\left(\!\!q_0x_+\!\!-\!j_xb\!+\!\frac{b}{2}\!\right)\!\!\right\}, \nonumber
\end{eqnarray}
where the function $\zeta_1(r)$ is defined after equation (\ref{Phip}).
For the ratio of the lattice to free space scattering amplitude we then obtain:
\begin{equation}\label{RlKP}
	\mathcal{R}_{l=1}=\frac{\eta^2\cot^2{\left({\eta}/{2}\right)}}{\left[1+{\sin\eta}/{\eta}\right]^{4}}\left[\frac{3}{2}+\frac{2\sin\eta}{\eta}+\frac{\sin{2\eta}}{4\eta}\right],
\end{equation}
and using Eq.~(\ref{eq:mKP}) the inverse BCS exponent in the lattice is expressed through the inverse BCS exponent in free space as
\begin{equation}\label{lambdacdelta}
\begin{split}
	\lambda_{c}^{\delta}&=\mathcal{R}_{l=1}\frac{m^*}{m}\lambda_{c}^{0}=\\
	&=\frac{\eta\cot(\eta/2)}{\left[1+{\sin\eta}/{\eta}\right]^{3}}\left[\frac{3}{2}+\frac{2\sin\eta}{\eta}+\frac{\sin{2\eta}}{4\eta}\right]\lambda_{c}^{0}.
\end{split}
\end{equation}

In the extreme limit of $G\gg1$ we have $\eta\simeq (\pi-2\pi/G)$, so that $\mathcal{R}_{l=1}\simeq\pi^4/G^2$ and $\lambda^{\delta}_c/\lambda_c^0\simeq\pi^2/G\ll{1}$.
We thus arrive at the same conclusion as in the previous section for sinusoidal lattices: in a very deep lattice the $p$-wave pairing of identical fermions is suppressed.
However, even for $G\simeq{20}$ the BCS exponent in the lattice exceeds the exponent in free space only by a factor of $1.7$ at the same density and short-range coupling strength 
(see Fig.~\ref{fig:KP}). 
It is thus crucial to understand what happens with the rates of inelastic decay processes in the lattice setup.

\begin{figure}[t]
\begin{centering}
\includegraphics[width=0.8\columnwidth]{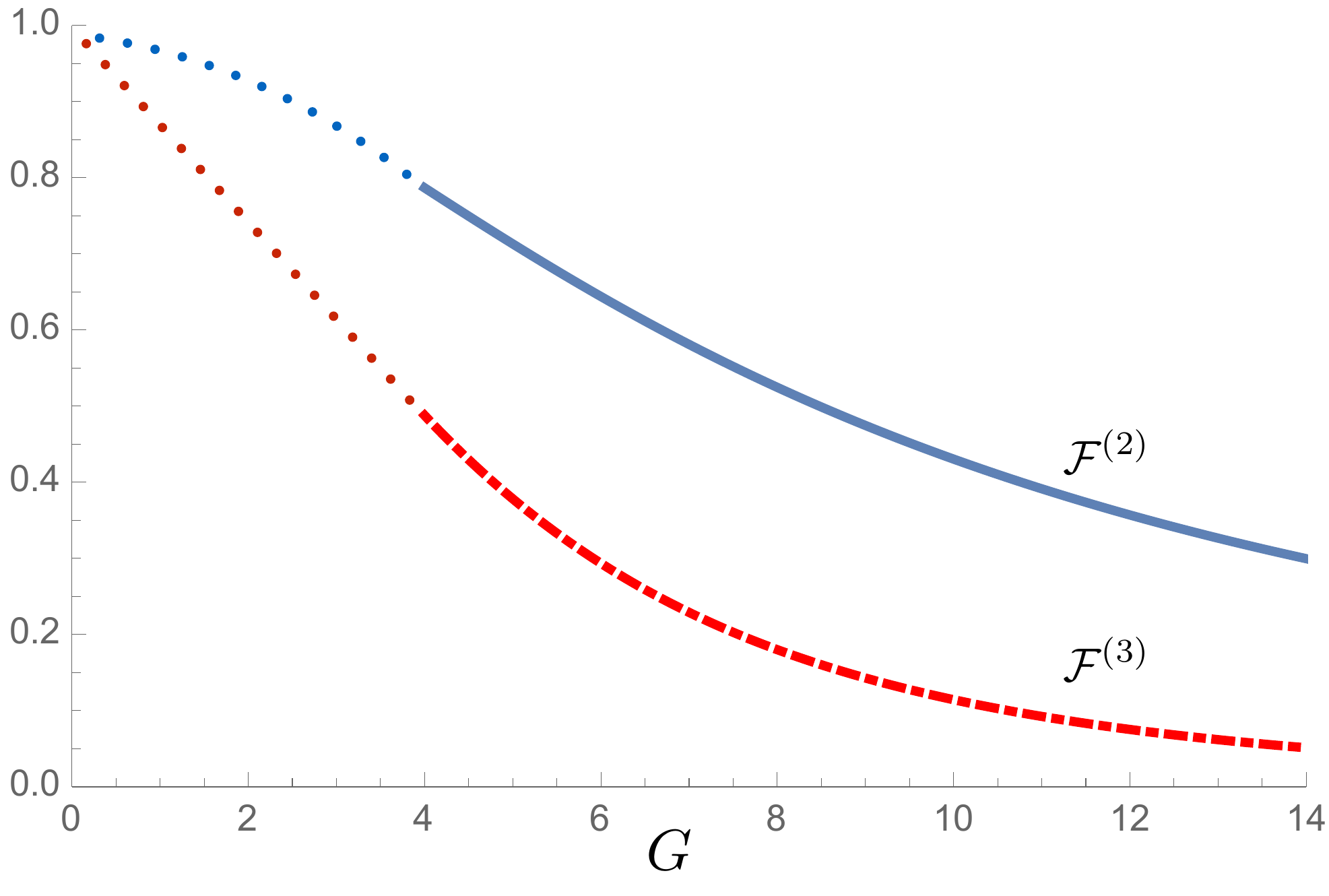}
\end{centering}
\vskip -4mm
\caption
{
Coefficients $\mathcal{F}^{(2)}$ and $\mathcal{F}^{(3)}$ as functions of the lattice depth $G$. The dotted parts of the curves show our expectation at $G\lesssim 1$, 
where the single-band approximation used in our calculations does not work.
}
\label{fig:decay}
\end{figure}

\section{Inelastic decay processes}\label{sec:inelastic}

We first consider the two-body relaxation, 
assuming that both colliding atoms are in an excited (internal energy $E_0$) hyperfine state and they relax to the ground state. 
The released hyperfine-state energy $2E_0$ goes to the kinetic energy of the atoms. 
It greatly exceeds the Fermi energy and the lattice potential depth, 
so that the relative motion of the atoms in the final state is described by a three-dimensional (3D) plane wave with a high momentum and they escape from the system. 
Then the number of relaxation events per unit time can be written in the form (see, e.g. Ref.~\cite{KSSh-88}):
\begin{equation}\label{w21}
	W_2=\int_{-\infty}^{\infty}dt\sum_i\rho_i\langle|\hat H'(0)\hat H'(t)|\rangle,
\end{equation}
where $\rho_i$ is the equilibrium density matrix, and $\hat H'$ is the Hamiltonian responsible for the relaxation process:
\begin{equation}\label{Hint}
	\hat H'(t)=\exp\{i\hat H_0t\}\hat H'(0)\exp\{-i\hat H_0t\},
\end{equation}
with $\hat H_0$ being the Hamiltonian of elastic interaction, and
\begin{equation}\label{Hin0}
\begin{split}
	\hat H'(0)=\int&{d\vec{\bf r}_1d\vec{\bf r}_2V_r(\vec{\bf r}_1-\vec{\bf r}_2)}\\
	&\times\left[\hat \Phi^{\dagger}(\vec{\bf r}_2)\hat\Phi^{\dagger}(\vec{\bf r}_1)\hat\psi(\vec{\bf r}_1)\hat\psi(\vec{\bf r}_2)+{\rm h.c.}\right].
\end{split}
\end{equation}
Here $\vec{\bf r}_1$ and $\vec{\bf r}_2$ are the 3D coordinates of the atoms, 
$\hat\psi(\vec{\bf r})$ is the field operator of the initial-state atoms,
 and $V_r(\vec{\bf r}_1-\vec{\bf r}_2)$ is the interaction potential causing the inelastic relaxation. 
The field operator of atoms in the final (ground) internal state is
\begin{equation}
	\Phi(\vec{\bf r})=\sum\nolimits_{\vec{\bf q}}\hat a_{\vec{\bf q}}\exp(i\vec{\bf q}\vec{\bf r}),
\end{equation}
and initially these states are not occupied. 
We thus have:
\begin{eqnarray}
&\!\!\!\!\!\!W_2=\int_{-\infty}^{\infty}dt\int d\vec{\bf r}_1d\vec{\bf r}_2d\vec{\bf r'}_1d\vec{\bf r'}_2V_r(\vec{\bf r}_1-\vec{\bf r}_2)V_r(\vec{\bf r'}_1-\vec{\bf r'}_2)  \nonumber \\
&\times\!\exp\{i\vec{\bf q}_1(\vec{\bf r}_1\!-\!\vec{\bf r'}_1\!)\!+\!i\vec{\bf q}_2(\vec{\bf r}_2\!-\!\vec{\bf r'}_2\!)\!-\!i[2E_0\!-\!(q_1^2\!+\!q_2^2)/2m]t\!\}    \nonumber \\
&\!\!\!\!\!\!\!\!\!\!\!\!\!\!\!\!\!\!\!\!\!\!\!\!\!\!\!\!\!\!\!\!\!\!\!\!\!\!\!\!\!\!\!\!\!\!\!\!\!\!\!\!\times\langle\hat\psi^{\dagger}(\vec{\bf r}_1,0)\hat\psi^{\dagger}(\vec{\bf r}_2,0)\hat\psi(\vec{\bf r'}_2,t)\hat\psi(\vec{\bf r'}_1,t)\rangle.     \label{w22}
\end{eqnarray}

The momenta $q_1$ and $q_2$ are large but the center of mass momentum $|\vec{\bf q}_1+\vec{\bf q}_2|$ is almost zero. 
The energy conservation law then reads:
\begin{equation}\label{econs}
	2E_0=\frac{p^2}{m},
\end{equation}
where $\vec{\bf p}=(\vec{\bf q}_1-\vec{\bf q}_2)/2$ is the relative momentum. 
From the summation over $\vec{\bf q}_1,\vec{\bf q}_2$ we turn to the integration over $\vec{\bf p}$ and $(\vec{\bf q}_1+\vec{\bf q}_2)$. 
The coordinate-dependent part of the exponent in Eq.~(\ref{w22}) takes the form:
\begin{equation}
\begin{split}
	&\exp\{i\vec{\bf q}_1(\vec{\bf r}_1-\vec{\bf r'}_1)+i\vec{\bf q}_2(\vec{\bf r}_2-\vec{\bf r'}_2)\}\\
	=&\exp\{i(\vec{\bf q}_1+\vec{\bf q}_2)(\vec{\bf R}-\vec{\bf R'})+i\vec{\bf p}(\vec{\bf r}-\vec{\bf r'})\},    
\end{split}
\end{equation}
where 
$\vec{\bf R}=(\vec{\bf r}_1+\vec{\bf r}_2)/2$, $\vec{\bf R'}=(\vec{\bf r'}_1+\vec{\bf r'}_2)/2$, $\vec{\bf r}=\vec{\bf r}_1-\vec{\bf r}_2$, and $\vec{\bf r'}=\vec{\bf r'}_1-\vec{\bf r'}_2$. 
The integration over $(\vec{\bf q}_1+\vec{\bf q}_2)$ yields:
\begin{equation}
\begin{split}
	&\int\frac{d(\vec{\bf q}_1+\vec{\bf q}_2)}{(2\pi)^3}\exp\{i(\vec{\bf q}_1+\vec{\bf q}_2)(\vec{\bf R}-\vec{\bf R'})\} \\
	&=\delta(\vec{\bf R}-\vec{\bf R'}),
\end{split}
\end{equation}
and the correlation function becomes:
\begin{eqnarray}
	&\langle\hat\psi^{\dagger}(\vec{\bf r}_1,0)\hat\psi^{\dagger}(\vec{\bf r}_2,0)\hat\psi(\vec{\bf r'}_2,t)\hat\psi(\vec{\bf r'}_1,t)\rangle= \\
	&\langle\hat\psi^{\dagger}(\vec{\bf R}+\vec{{\bf r}\,'}/2)\hat\psi^{\dagger}(\vec{\bf R}-\vec{{\bf r}\,'}/2)\hat\psi(\vec{\bf R}-\vec{\bf r}/2,t)\hat\psi(\vec{\bf R}+\vec{\bf r}/2,t)\rangle.  \nonumber
\end{eqnarray}
Characteristic times $t$ on which the correlation function changes are of the order of the inverse Fermi energy or even larger. 
They are much longer than the times $\sim E_0^{-1}$ that dominate the integral over $dt$ in Eq.~(\ref{w22}). 
Therefore, we may put $t=0$ in the correlation function, which reduces Eq.~(\ref{w22}) to
\begin{eqnarray}\label{w23}
	&W_2=\int \tilde W_2(\vec{\bf r},\vec{\bf r'})d\vec{\bf R}d\vec{\bf r}d\vec{\bf r'}   \nonumber \\
	&\!\!\!\!\!\!\!\!\!\times\!\langle\hat\psi^{\dagger}(\vec{\bf R}\!+\!\vec{\bf r}'/2)
	\hat\psi^{\dagger}(\vec{\bf R}\!-\!\vec{\bf r}'/2)
	\hat\psi(\vec{\bf R}\!-\!\vec{\bf r}/2)\hat\psi(\vec{\bf R}\!+\!\vec{\bf r}/2)\rangle,
\end{eqnarray}
with
\begin{equation}\label{w21tilde}
	\!\tilde W_2\!=\!\!\!\int\!\! V_r(\vec{\bf r})V_r(\vec{\bf r'})\exp\{i\vec{\bf p}(\vec{\bf r}\!-\!\vec{\bf r'})\}\delta\!\left(\!2E_0\!-\!\frac{p^2}{m}\!\right)\!\frac{d\vec{\bf p}}{(2\pi)^2}.\!\!\!
\end{equation}

In the quasi-2D geometry the field operator can be written as 
$\hat\psi(\vec{\bf r}_{1,2})=\psi_0(z_{1,2})\psi({\bf r}_{1,2})$, 
where ${\bf r}_{1,2}$ is the 2D vector in the $x,y$ plane, 
and
\begin{equation}
	\psi_0(z_{1,2})=\frac{1}{(\pi l_0^2)^{1/4}}\exp\left[-\frac{z_{1,2}^2}{2l_0^2}\right]
\end{equation}
is the wavefunction in the tightly confined $z$-direction \cite{Note}. 
As the inelastic relaxation occurs at interparticle distances much smaller than the confinement length $l_0$, 
the product of four field operators in Eq.~(\ref{w23}) becomes
\begin{equation}
\begin{split}
	\hat\psi^{\dagger}({\bf R}+{\bf r}'/2)&\hat\psi^{\dagger}({\bf R}\!-{\bf r}'/2)\\
	&\times\hat\psi({\bf R}\!-{\bf r}/2)\hat\psi({\bf R}\!+{\bf r}/2)\psi_0^4(Z),
\end{split}	
\end{equation}
where ${\bf R},{\bf r}$ and ${\bf r'}$ are 2D vectors in the $x,y$ plane, and $Z=(z_1+z_2)/2$. Integrating over $Z$ in Eq.~(\ref{w23}) we then have:
\begin{eqnarray}  \label{w24}
	&W_2=\int \tilde w_2({\bf r},{\bf r'})d{\bf R}d{\bf r}d{\bf r'}   \nonumber \\
	&\!\!\!\!\!\!\!\!\!\times\!\langle\hat\psi^{\dagger}({\bf R}\!+\!{\bf r}'/2)
	\hat\psi^{\dagger}({\bf R}\!-\!{\bf r}'/2)
	\hat\psi({\bf R}\!-\!{\bf r}/2)\hat\psi({\bf R}\!+\!{\bf r}/2)\rangle,
\end{eqnarray}
where
\begin{equation}\label{w22tilde}
	\tilde w_2({\bf r},{\bf r'})=\int \tilde W_2(\vec{\bf r},\vec{\bf r'})\frac{dzdz'}{\sqrt{2\pi}l_0},
\end{equation}
 with $z=z_1-z_2$ and $z'=z_1'-z_2'$.

Using expansion (\ref{Psi-expansion}) one can express the averaged product of four 2D field operators in terms of the standard Slater determinants
$\mathcal{D}({\bf r}, {\bf R}; {\bf k}_1, {\bf k}_2 )$:
\begin{eqnarray}\label{4-average}
	&&
	\langle\hat\psi^{\dagger}({\bf R}+{\bf r}'/2)
	\hat\psi^{\dagger}({\bf R}-{\bf r}'/2)
	\hat\psi^{}({\bf R}-{\bf r}/2)\hat\psi^{}({\bf R}+{\bf r}/2)\rangle
	\nonumber \\
	&&
	= 
	\frac{1}{2!} \sum_{{\bf k}_1, {\bf k}_2}N_{k_1}N_{k_2}
	\mathcal{D}^*({\bf r}', {\bf R} ; {\bf k}_1, {\bf k}_2 )
	\mathcal{D}({\bf r}, {\bf R} ; {\bf k}_1, {\bf k}_2 ),
\end{eqnarray}
where $N_{k}$ is the Fermi distribution function, and
\begin{equation}\label{Slater}
\begin{split}
	\mathcal{D}&({\bf r}, {\bf R} ; {\bf k}_1, {\bf k}_2 ) \\
	=&\mathrm{Det}
\left(
  \begin{array}{cc}
    \chi_{{\bf k}_1}({\bf R} +{\bf r}/2) & \chi_{{\bf k}_1}({\bf R} - {\bf r}/2) \\
    \chi_{{\bf k}_2}({\bf R} +{\bf r}/2) & \chi_{{\bf k}_2}({\bf R} -{\bf r}/2) \\
  \end{array}
\right).
\end{split}
\end{equation}
The distance ${\bf r}$ between relaxing particles is small compared to the lattice period and particle wavelengths.
Therefore, all the wavefunctions entering Eq.~(\ref{Slater}) should be taken within the same lattice cell $(n,m)$ of the considered 2D lattice, so that
Eq.~(\ref{w24}) will contain only one double lattice summation over $n$ and $m$. 
The Slater determinant (\ref{Slater}) within a given cell $(n,m)$ contains a factor $\exp{[i(k_{1x}+k_{2x})bn + i(k_{1y}+k_{2y})bm]}$ [see Eq.~(\ref{eq:chi-exact})], 
which does not contribute to the product $\mathcal{D}^*\mathcal{D}$.
Below we will imply that the corresponding exponential factors have been already extracted from the wavefunctions. In the leading (linear) order in small ${\bf r}$ we have:
\begin{equation}\label{expansion}
	\chi_{{\bf k}}\left({\bf R}\pm \frac{{\bf r}}{2}\right) \approx \chi_{{\bf k}}({\bf R})\left\{1
	\pm
	\frac{1}{2}{\bf r} \cdot {\bf \nabla}_{\bf R} \ln{[\chi_{{\bf k}}({\bf R})]}\right\},
\end{equation}
where the Slater determinant takes the form:
\begin{equation}\label{Slater-expansion}
\begin{split}
	\mathcal{D}&({\bf r}, {\bf R}; {\bf k}_1, {\bf k}_2)=
	\chi_{{\bf k}_1}({\bf R}) \chi_{{\bf k}_2}({\bf R})\\
	&\times {\bf r}\cdot {\bf \nabla}_{\bf R}\left\{\ln{[\chi_{{\bf k}_1}({\bf R})]}
	- \ln{[\chi_{{\bf k}_2}({\bf R})]} \right\}.
\end{split}
\end{equation}
As the leading contribution to the scattering of slow identical fermions comes from the
$p$-wave scattering channel, the expression in the curly
brackets in Eq.~(\ref{Slater-expansion}) is linear (in the leading order) in
the difference $({\bf k}_1 - {\bf k}_2)$. For instance, the ``$x$-component''
of this expression has the form:
\begin{eqnarray}\label{Slater-expansion-x}
	&& \ln{[\chi_{k_{1x}}(X)]}
	- \ln{[\chi_{k_{2x}}(X)]}  = \nonumber \\
	&& \frac{i(k_{1x}-k_{2x})b}{2\sin{(\eta/2)}}
	\frac{\sin{[q(X-nb)]}}{\cos{[q(X-nb) + b/2)]}},
\end{eqnarray}
where $X$ varies from $(n-1)b$ to $nb$ (see Eq.~(\ref{eq:chi-exact})).
In the considered low density limit ($kb \ll 1$) we may put ${\bf k}_1 = {\bf k}_2=0$ in the product
$\chi_{{\bf k}_1}({\bf r}_1) \chi_{{\bf k}_2}({\bf r}_1)$ in Eq.~(\ref{Slater-expansion}). 
As a result we transform Eq.~(\ref{w24}) to
\begin{equation}\label{w25}
\begin{split}
	W_2\!=\!\frac{\mathcal{F}_2(\eta)}{2}&\int\!\! {\bf r}{\bf r'}w_2({\bf r},{\bf r'})d{\bf r}d{\bf r'}\\
	&\times\int\!\! N_{k_1}N_{k_2}(k_1^2\!+\!k_2^2)\frac{d{\bf k}_1d{\bf k}_2}{(2\pi)^4}.
\end{split}
\end{equation}
The quantity $\mathcal{F}_2 $ is determined by the integral over the 2D lattice:
\begin{eqnarray}\label{F-cal}
\mathcal{F}_2 =
	&& \frac{1}{8\sin^2{\eta/2}} \sum^{\infty}_{n, m} \int^{\infty}_{-\infty}
	\int^{\infty}_{-\infty} dx dy A_{n}(x)A_{m}(y) \nonumber \\
	&&  \times |\chi_{0}(x)|^4 |\chi_{0}(y)|^4 \left[P^2(x) + P^2(y)\right],
\end{eqnarray}
where the functions $A_{j}(x)$ are defined below Eq.~(\ref{eq:chi-exact}),
and the function $P$ results from the differentiation of the curly brackets in Eq.~(\ref{Slater-expansion})
with the use of Eq.~(\ref{Slater-expansion-x}):
\begin{eqnarray}\label{P(u)}
	P(u) \equiv \frac{d}{du} \, \frac{\sin{[qu]}}{\cos{[q(u + b/2)]}}.
\end{eqnarray}
Performing the integration in Eq.~(\ref{F-cal}) we find
\begin{equation}     \label{FR}
	\mathcal{F}_2(\eta)=\mathcal{R}_{l=1}(\eta),
\end{equation}
with the lattice factor $\mathcal{R}_{l=1}$ given by Eq.~(\ref{RlKP}).

In the absence of the 2D lattice (i.e. in free 2D space) we also arrive at Eq.~(\ref{w24}).
Then, using $\chi_{{\bf k}}({\bf r}) =\exp{(i{\bf k}{\bf r})}$, the Slater determinant becomes:
\begin{eqnarray}\label{Slater-free}
	\mathcal{D}({\bf r}, {\bf R}; {\bf k}_1, {\bf k}_2 ) 
	\simeq 
	i({\bf k}_{1}-{\bf k}_{2}){\bf r}\exp\left[{i({\bf k}_1 + {\bf k}_2){\bf R}}\right].
\end{eqnarray}
Performing integrations we get Eq.~(\ref{w25}) with $\mathcal{F}_2$ replaced by unity.
Thus, we obtain that in the lattice the two-body inelastic relaxation is reduced by a factor of $\mathcal{F}_2$ compared to free space:
\begin{equation}\label{w2fin}
	W_2^{\rm lat}=\mathcal{F}_2(\eta)W_2^{\rm free}.
\end{equation}
The function $\mathcal{F}_2(\eta)$ following from Eqs.~(\ref{RlKP}) and (\ref{FR}) is displayed in Fig.~4 versus the lattice depth $G$, which is related to $\eta$ by Eq.~(\ref{eq:eta}).

We complete this section with the discussion of three-body recombination, 
assuming that the binding energy of the molecule formed in this process greatly exceeds the Fermi energy and the lattice depth. 
In this case the kinetic energies of the molecule and atom in the output channel of the recombination are very high and they escape from the system. 
The results for the ratio of the three-body recombination rate in the lattice to the rate in free space are obtained in a way similar to that for the two-body relaxation. 
The number of recombination events per unit time, $W_3$, is given by Eq.~(\ref{w21}) in which $\hat H'(t)$ follows from Eq.~(\ref{Hint}), 
and the Hamiltonian $\hat H'(0)$ is given by
\begin{equation}\label{Hin03}
\begin{split}
	\hat H'(0)&=\int{d \vec{\bf r}_1 d \vec{\bf r}_2 \vec{\bf r}_3}
	V(\vec{\bf r}_1, \vec{\bf r}_2, \vec{\bf r}_3)\times \\
	&\left[\hat{B}^{\dag}(\vec{\bf r}_1, \vec{\bf r}_2) \hat{\psi}^{\dag}(\vec{\bf r}_3)\hat{\psi}(\vec{\bf r}_3)\hat{\psi}(\vec{\bf r}_2)\hat{\psi}(\vec{\bf r}_1) + \mathrm{h.c.}
	\right]
\end{split}
\end{equation}
with $V(\vec{\bf r}_1, \vec{\bf r}_2, \vec{\bf r}_3)$ being the sum of three pair interaction potentials,
and $\hat B^{\dagger}(\vec{\bf r}_1,\vec{\bf r}_2)$ the field operator of the molecules. 
The latter can be written as
\begin{equation}\label{B-operator}
	\hat{B}^{\dag}(\vec{\bf r}_1, \vec{\bf r}_2) =
	\sum\nolimits_{\vec{\bf q}, s} \exp^{-i\vec{\bf q}\vec{\bf R}}
	\chi^*_{s}(\vec{\bf r})
	\hat{b}^{\dag}_{\vec{\bf q} s},
\end{equation}
where $\hat{b}^{\dag}_{\vec{\bf q} s}$ is the creation operator of the molecule with momentum $\vec{\bf q}$ in the internal state $s$, 
$\chi_{s}(\vec{\bf r})$ is the wavefunction of this state, 
and the notations for coordinates are the same as in the above discussion of two-body relaxation.

Initially molecules are not present in the system and, hence, for the average of the molecular field operators we have:
\begin{equation}
\begin{split}
	&\langle\hat{B}(\vec{{\bf r}}\,'_1,\vec{{\bf r}}\,'_2,0)\hat{B}^{\dag}(\vec{\bf r}_1,\vec{\bf r}_2, t)\rangle
	=\chi_{s}(\vec{\bf r}\,')\chi^*_{s}(\vec{\bf r}) \\
	&\times\sum_{\vec{\bf q}, s} \exp\left\{-i\vec{\bf q}(\vec{\bf R}-\vec{\bf R}')+i\left(\frac{q^2}{4m}-E_s\right)t\right\},
\end{split}
\end{equation}
with $E_s$ being the binding energy of the molecule in the state $s$;
$\vec{\bf R}=(\vec{\bf r}_1 + \vec{\bf r}_2)/2$;
$\vec{\bf r}=\vec{\bf r}_1 - \vec{\bf r}_2$
(and similarly for $\vec{\bf R}'$ and $\vec{\bf r}\,'$). 
The momentum $p$ of the atom in the outgoing recombination channel is very high, and the states with such momenta are not initially occupied. 
Therefore, we get
\begin{equation}
\begin{split}
	\langle\hat{\psi}({\vec{\bf r}}\,'_3, 0) &\hat{\psi}^{\dag}(\vec{\bf r}_3, t)\rangle \\
	&=\sum_{\vec{\bf p}}\exp{\left\{-i\vec{\bf p}(\vec{\bf r}_3 -\vec{{\bf r}}\,'_3)+i\frac{p^2}{2m}t\right\}}.  
\end{split}
\end{equation}
Thus, the initial expression for $W_3$ (Eq.~(\ref{w21}) with $\hat H'(t)$ (\ref{Hint}) and $\hat H'(0)$ (\ref{Hin03})) takes the form:
\begin{eqnarray}\label{w31}
	&&
	W_3=\int^{\infty}_{-\infty} dt\int
	d\vec{\bf R}d\vec{\bf R}'d\vec{\bf r}d\vec{\bf r}'d\vec{\bf u}d\vec{\bf u}'
	V(\vec{\bf r}', \vec{\bf u}')V(\vec{\bf r}, \vec{\bf u})
	\nonumber \\
	&&
	\sum_{\vec{\bf p}, \vec{\bf q},s}
	\exp{\left\{-i\left[\vec{\bf p}(\vec{\bf R}-\vec{\bf R}' + \vec{\bf u}-\vec{\bf u}')
	+\vec{\bf q}(\vec{\bf R}- \vec{\bf R}')\right]\right\}}    \nonumber \\
	&&\times\chi_{s}(\vec{\bf r}')\chi^*_{s}(\vec{\bf r})\exp{\left[i\left(\frac{p^2}{2m} + \frac{q^2}{4m} - E_s\right)t\right]}
	\nonumber \\
	&& 
	\times \langle \hat\psi^{\dagger}(\vec{\bf R}' + \vec{\bf r}'/2, 0)
	\hat\psi^{\dagger}(\vec{\bf R}' - \vec{\bf r}'/2, 0)
	\hat\psi^{\dagger}(\vec{\bf R}' + \vec{\bf u}', 0)
	\nonumber \\
	&&
	\times\hat\psi^{\dagger}(\vec{\bf R} + \vec{\bf u}, t)
	\hat\psi^{}(\vec{\bf R} - \vec{\bf r}/2, t)
	\hat\psi^{}(\vec{\bf R} + \vec{\bf r}/2, t)\rangle,
\end{eqnarray}
where $V(\vec{{\bf r}\,'}, \vec{{\bf u}\,'})\equiv{V(\vec{\bf R}+\vec{\bf r}/2,\vec{\bf R}-\vec{\bf r}/2,\vec{\bf R} +\vec{\bf u})}$ and $\vec{\bf u}=\vec{\bf r}_3-\vec{\bf R}$.
Omitting a small difference between $q$ and $p$ in the time-dependent exponent transforms it to $\exp{[i(3p^2/4m-E_s)t]}$ 
and after putting $t=0$ in the correlation function the integration over $t$ yields $\delta(3p^2/4m-E_s)$. 
The summation over $\vec{\bf q}$ gives $\delta(\vec{\bf R}-\vec{\bf R}')$. 
As a result, Eq.~(\ref{w31}) reduces to
\begin{equation}\label{w32}
\begin{split}
	W_3=&\int\tilde W_3(\vec{\bf r},\vec{\bf r}',\vec{\bf u},\vec{\bf u}')d\vec{\bf R}d\vec{\bf r}d\vec{\bf r}'d\vec{\bf u}d\vec{\bf u}'\times\\
	&\langle\hat\psi^{\dagger}(\vec{\bf R}+\vec{\bf r}'/2)\hat\psi^{\dagger}(\vec{\bf R}-\vec{\bf r}'/2)\hat\psi^{\dagger}(\vec{\bf R}+\vec{\bf u}')\times \\
	&\hat\psi(\vec{\bf R}+\vec{\bf u})\hat\psi^{}(\vec{\bf R}-\vec{\bf r}/2)\hat\psi^{}(\vec{\bf R}+\vec{\bf r}/2)\rangle,
\end{split}
\end{equation}
with
\begin{eqnarray}\label{tildew31}
	&&\tilde{W}_3(\vec{\bf r},\vec{\bf r}',\vec{\bf u},\vec{\bf u}')=V(\vec{\bf r}, \vec{{\bf u}})V(\vec{\bf r}', \vec{\bf u}' )\int \frac{d\vec{\bf p}}{(2\pi)^2} 
	\nonumber  \\
	&&\times\exp\left[{i\vec{\bf p}(\vec{\bf u}-\vec{\bf u}')}\right]  \sum_{s}\delta\left(\frac{3p^2}{4m}{-}E_s\right)\chi^*_s(\vec{\bf r})\chi_s(\vec{\bf r}').
\end{eqnarray}
Integrating out the motion of particles in the tightly confined $z$-direction in a way similar to that for the two-body relaxation, we transform Eq.~(\ref{w32}) to
\begin{eqnarray}\label{w33}
	&&\!\!W_3\!=\!\!\!\int\!\!\! d{\bf R}d{\bf r}d{\bf r}'d{\bf u}d{\bf u}' \tilde w_3({\bf r},{\bf r}'\!,{\bf u},{\bf u}')\langle\hat\psi^{\dagger}({\bf R}\!+\!{\bf r}'\!/2)\hat\psi^{\dagger}({\bf R}\!-\!{\bf r}'\!/2)     	
	\nonumber  \\
	&&\times\hat\psi^{\dagger}({\bf R}+{\bf u}')\hat\psi({\bf R}+{\bf u})\hat\psi({\bf R}-{\bf r}/2)\hat\psi({\bf R}+{\bf r}/2)\rangle,
\end{eqnarray}
where ${\bf R},{\bf r},{\bf u}$ and ${\bf R}',{\bf r}',{\bf u}'$ are 2D vectors in the $x,y$ plane and
\begin{eqnarray}\label{tildew32}
	\tilde w_3({\bf r},{\bf r}',{\bf u},{\bf u}')=\int \tilde W_3(\vec{\bf r},\vec{\bf r}',\vec{\bf u},\vec{\bf u}')\frac{dzdz'du_zdu'_z}{\sqrt{3}\pi l_0^2}.
\end{eqnarray}

Similarly to Eq.~(\ref{4-average}), the averaged product of six fermionic field operators is represented as
\begin{eqnarray}\label{6-average}
	&& S_3 \equiv {\langle{
	\hat\psi^{\dagger}({\bf r}'_1)\hat\psi^{\dagger}({\bf r}'_2)
	\hat\psi^{\dagger}({\bf r}'_3)
	\hat\psi^{}({\bf r}_3)\hat\psi^{}({\bf r}_2)\hat\psi^{}({\bf r}_1)}
	\rangle}  \nonumber \\
	&&=\frac{1}{3!} \sum_{{\bf k}_1, {\bf k}_2,{\bf k}_1}  N_{k_1}N_{k_2}N_{k_3}
	 \\
	&&\times\mathcal{D}^*({\bf r}'_1, {\bf r}'_2, {\bf r}'_3  ; {\bf k}_1, {\bf k}_2, {\bf k}_3 )\mathcal{D}({\bf r}_1, {\bf r}_2, {\bf r}_3  ; {\bf k}_1, {\bf k}_2, {\bf k}_3) \nonumber,
\end{eqnarray}
where $\mathcal{D}({\bf r}_1, {\bf r}_2, {\bf r}_3  ; {\bf k}_1, {\bf k}_2, {\bf k}_3 )$
is the (Slater) determinant of the $3 \times 3$ matrix $\{\chi_{{\bf k}_i}({\bf r}_j)\}$.
Using the expansion of the wavefunctions in (small) relative coordinates
${\bf r} = {\bf r}_{1} - {\bf r}_{2}$ and
${\bf u} = {\bf r}_{3} - ({\bf r}_{1} + {\bf r}_{2})/2$ we find that
$\mathcal{D}$ is bilinear in the components of these quantities:
\begin{eqnarray}\label{D-expansion}
	&&\!\! \mathcal{D}({\bf R}\!+\!{\bf r}/2, {\bf R}\!-\!{\bf r}/2,
	{\bf R}\!+\!{\bf u} ; {\bf k}_1, {\bf k}_2, {\bf k}_3 ) \nonumber \\
	&&\!=\! \frac{1}{2}\chi_{{\bf k}_1}({\bf R})\chi_{{\bf k}_2}({\bf R})\chi_{{\bf k}_3}({\bf R})\sum_{\alpha, \beta}(r_{\alpha}u_{\beta}-r_{\beta}u_{\alpha}) \nonumber \\
	&&\times\{\nabla_{\alpha}[ \ln{\chi_{{\bf k}_1}({\bf R})}-\ln{\chi_{{\bf k}_2}({\bf R})}] 
	\nabla_{\beta}[ \ln{\chi_{{\bf k}_2}({\bf R})}-\ln{\chi_{{\bf k}_3}({\bf R})}] \nonumber \\
	&&\!-\!\nabla_{\beta}[ \ln{\!\chi_{{\bf k}_1\!}({\bf R})}\!-\!\ln{\!\chi_{{\bf k}_2\!}({\bf R})}]\nabla_{\alpha}[ \ln{\!\chi_{{\bf k}_2\!}({\bf R})}\!-\!\ln{\!\chi_{{\bf k}_3\!}({\bf R})}]\},
\end{eqnarray}
where $\alpha,\beta=\{x,y\}$.
Using Eqs. (\ref{Slater-expansion-x}) and (\ref{P(u)}), in the leading
order in small relative wavevectors
equation (\ref{D-expansion}) takes the form:
\begin{eqnarray}\label{D-expansion-approx}
	&& \mathcal{D}({\bf R}+{\bf r}/2, {\bf R}-{\bf r}/2,
	{\bf R}+{\bf u} ; {\bf k}_1, {\bf k}_2, {\bf k}_3 ) \simeq
	\frac{[\chi_{0}({\bf R})]^3 b^2}{4\sin^2{(\eta/2)}}\nonumber \\
	&&\!\!\!\!\times\sum_{\alpha, \beta}(r_{\alpha}u_{\beta}\!-\!r_{\beta}u_{\alpha})
	(k_1\! -\!k_2)_{\alpha}(k_3\! -\! k_2)_{\beta}
	P(R_{\alpha})P(R_{\beta}).
\end{eqnarray}
Substituting the result of Eq.~(\ref{D-expansion-approx}) into equation (\ref{6-average}) we find for the correlation function: 
\begin{eqnarray}\label{S-3}
	&&S_3 =
	\frac{|\chi_{0}({\bf R})|^6 b^4}{2^7\sin^4{(\eta/2)}} 
	\nonumber \\
	&&
	\times
	\frac{1}{3}\int N_{k_1} N_{k_2}N_{k_3}[k^2_1k^2_2 + k^2_1k^2_3 + k^2_1k^2_3]
	\frac{d^2k_1d^2k_2d^2k_3}{(2\pi)^6}
	\nonumber \\
	&&
	\times \sum_{\alpha, \beta}(r_{\alpha}u_{\beta}-r_{\beta}u_{\alpha})
	(r'_{\alpha}u'_{\beta}-r'_{\beta}u'_{\alpha})
	P^2(R_{\alpha})P^2(R_{\beta}).
\end{eqnarray}
Having in mind that only the terms with $\beta \neq \alpha$ contribute to the summation over
2D Cartesian indices, from Eqs.~(\ref{w33}), (\ref{6-average}), and (\ref{S-3}) we obtain for the decay rate:
\begin{eqnarray}\label{w33-final}
&& W_3=\frac{\mathcal{F}_3(\eta)}{12}\int d{\bf r}d{\bf r}'d{\bf u}d{\bf u}' \tilde w_3({\bf r},{\bf r}',{\bf u},{\bf u}')
[\vec{\bf r} \times \vec{\bf u}]_z [\vec{\bf r}' \times \vec{\bf u}']_z
\nonumber  \\
&& 
\times\!\!\int\!\!\! N_{k_1} N_{k_2}N_{k_3}[k^2_1k^2_2 + k^2_2k^2_3 + k^2_1k^2_3]
\frac{d^2k_1d^2k_2d^2k_3}{(2\pi)^6},
\end{eqnarray}
where we expressed the combination $r_1u_2 - r_2u_1$ in terms of
3D vectors $\vec{\bf r}$ and $\vec{\bf u}$. 
The quantity $\mathcal{F}_3(\eta)$ in Eq.~(\ref{w33-final})
is given by 
\begin{eqnarray}    \label{F3-def}
\!\!\mathcal{F}_3(\eta)\!=\! \frac{b^2}{16\sin^4{(\eta/2)}}
\!\int\!\! d{\bf R} |\chi_0({\bf R})|^6 P^2(X)P^2(Y).
\end{eqnarray}
Here the integration over ${\bf R}$ is only in the 2D lattice cell, while the summation over all lattice 
cells resulted in the multiplication of the result by the cell number $1/b^2$.
Performing the integration we obtain:
\begin{equation}    \label{F3-value}
\mathcal{F}_3(\eta)=
\eta^4\cot^4\left(\eta/2\right)\left[1+\frac{\sin\eta}{\eta}\right]^{-4}.
\end{equation}

Let us now compare the result of Eq.~(\ref{w33-final}) with that in free space. 
Taking the wavefunction $\chi_{\bf k} =\exp(i{\bf kr})$ the expression for $\mathcal{D}$ becomes:
\begin{equation}\label{D-expansion-approx-free}
\begin{split}
	\mathcal{D}&({\bf R}\!+\!{\bf r}/2, {\bf R}\!-\!{\bf r}/2,{\bf R}\!+\!{\bf u} ;\! {\bf k}_1, {\bf k}_2, {\bf k}_3)\\
	&=\!\exp[i({\bf k}_1\!\!+\! {\bf k}_2\!+\! {\bf k}_3 ){\bf R}] \\
	&\times i\sum_{\alpha, \beta}(r_{\alpha}u_{\beta}-r_{\beta}u_{\alpha})
	(k_1 -k_2)_{\alpha}(k_3 - k_2)_{\beta}.
\end{split}
\end{equation}
Using $\mathcal{D}$ (\ref{D-expansion-approx-free}) in Eqs.~(\ref{6-average}) and (\ref{w33}) 
we arrive at the recombination rate given by Eq.~(\ref{w33-final})  without the factor $\mathcal{F}_3$ in the right hand side. 
Thus, the relation between the recombination decay  rate in free space and the one in the 2D lattice reads:
\begin{equation} \label{w3fin}
	W_3^{\rm lat}=W_3^{\rm free}\mathcal{F}_3(\eta).
\end{equation}
The function $\mathcal{F}_3$ is shown in Fig.~4 versus the lattice depth $G$ related to $\eta$ by Eq.~(\ref{eq:eta}). 
The results obtained in this section indicate that both two-body and three-body inelastic collisions are significantly suppressed in the lattice setup even at moderate depths.

For usual sinusoidal optical lattices used in experiments with ultracold atoms, one can proceed along the same lines as in the case of the 2D Kronig-Penney model. 
For the two-body relaxation, Eqs.~(\ref{w21})-(\ref{Slater}) remain the same.  
Then, for fairly deep lattices ($b/\xi_0\gtrsim 4$) where the function $\chi_{\bf k}({\bf r})$ can be still used in the form (\ref{eq:wavefunction}), 
we obtain the ratio of the lattice to free space relaxation rate $W_2^{\rm sl}/W_2^{\rm free}\simeq\mathcal{R}_{l=1}$, with the factor $\mathcal{R}_{l=1}$ given by Eq.~(\ref{Rl1}). 
The calculations for the three-body recombination are more involved. 
The estimate using an analogy with the Kronig-Penney model at large $G$, leads to the ratio of the lattice to free space recombination rate $W_3^{\rm sl}/W_3^{\rm free}{\sim}\mathcal{R}_{l=1}^2$. 
In particular, for $b/\xi_0=4$ ($m^*/m\simeq 5$) the two-body relaxation is suppressed by a factor of $5$ and the three-body recombination by about a factor of 25.

\section{Conclusions and outlook}\label{sec:conclusion}

The results of the present paper indicate that there are possibilities to create the superfluid topological $p_x+ip_y$ phase of atomic lattice fermions. 
In deep lattices the $p$-wave superfluid pairing is suppressed 
and even for moderate lattice depths the BCS exponent is larger than in free space at the same density and short-range coupling strength.
However, the lattice setup significantly reduces the inelastic collisional losses, 
so that one can get closer to the $p$-wave Feshbach resonance and increase the interaction strength without inducing a rapid decay of the system. 

For ultracold $^6$Li the $p$-wave resonance is observed for atoms in the lowest hyperfine state $(1/2,1/2)$ \cite{Salomon,Ketterle3,Salomon2,Ticknor,Ueda,Nakasuji}, 
and the only decay channel is three-body recombination. 
In the 2D Kronig-Penney lattice with the depth $G\simeq 12$ and the period $b\simeq 200$ nm ($m^*/m\simeq 2$ and $\mathcal{R}_{l=1} m^*/m\approx 0.7$), 
at $k_Fb\simeq 0.5$ the Fermi energy is close to 100 nK and the 2D density is about $0.5\times 10^8$ cm$^{-2}$. 
Slightly away from the Feshbach resonance (at the scattering volume $V_{sc}\simeq{8}\times{10^{-15}}$ cm$^3$) we are still in the weakly interacting regime, 
and the 3D recombination rate constant is $\alpha_{rec}^{3D}\sim 10^{-24}$ cm$^6$/s~\cite{Salomon}. 
Then, using Eq.~(\ref{eq:Tc0}) and the quasi2D scattering amplitude expressed through $V_{sc}$ and the tight confinement length $l_0=\sqrt{1/m\omega_0}$~\cite{Pricoupenko},
 for the confinement frequency $\omega_0{\simeq}100$ kHz we obtain the BCS critical temperature $T_c\simeq 5$ nK. 
 The 2D recombination rate constant is $\alpha_{rec}^{2D}\approx \mathcal{F}_3\alpha_{rec}^{3D}/\sqrt{3}\pi l_0^2$ 
 and with $\mathcal{F}_3\simeq 0.05$ at $G\simeq 12$ we arrive at the decay time $\tau_{rec}\sim 1/\alpha_{rec}^{2D}n^2$ approaching 1 second.

The $p$-wave Feshbach resonance for $^{40}$K occurs between atoms in the excited hyperfine state $(9/2,-7/2)$. 
Therefore, there is also a decay due to two-body relaxation. 
For the same parameters as in the discussed Li case ($G, V_{sc}, l_0, b, n$) we then have the Fermi energy $E_F\simeq 20$ nK, and the BCS transition temperature approaches 1 nK. 
Using experimental values for the relaxation and recombination rate constants in 3D~\cite{Jin} and retransforming them to the 2D lattice case, 
we obtain the relaxation and recombination times of the order of seconds.

It is worth mentioning that  in recently proposed subwavelength lattices~\cite{Dalibard,Cirac,Kimble} one may have the lattice period $b\simeq{60}$ nm, 
and for the same $k_Fb\simeq{0.5}$ the density and Fermi energy will be higher by an order of magnitude. 
Then departing further from the Feshbach resonance one gets the same BCS exponent as above, and the critical temperature for $^6$Li will be $\sim 50$ nK. 
The recombination time is again on the level of a second.

\section*{Acknowledgments}

We are grateful to D.V. Kurlov and D.S. Petrov for useful comments. 
This research was supported in part by the National Science Foundation under Grant No. PHY11-25915.
G.V.S. is grateful to the Kavli Institute of Theoretical Physics at Santa Barbara for hospitality during the workshop on Universality of Few-Body Systems (November-December, 2016), 
where part of the work has been done. 
The research leading to these results has received funding from the European Research Council under European Community's Seventh Framework Programme 
(FP7/2007-2013 Grant Agreement no. 341197).
The work was supported in part by the RFBR Grant 17-08-00742 (A.F.) and by the Basic research program of HSE (V.Y.).

\end{document}